# Influence of Dimensionality on the Charge Density Wave Phase of 2H-TaSe$_2$

*Sugata Chowdhury*[*], *Albert F. Rigosi, Heather M. Hill, Andrew Briggs, David B. Newell, Helmuth Berger, Angela R. Hight Walker, and Francesca Tavazza*

Dr. S. Chowdhury
Department of Physics and Astronomy, Howard University, Washington, DC 20059, United States
IBM-HBCU Quantum Center, Howard University, Washington, DC 20059, USA
National Institute of Standards and Technology, Gaithersburg, MD 20899, United States
Email: sugata.chowdhury@howard.edu; sugata.chowdhury@nist.gov

Dr. A. F. Rigosi, Dr. H. M. Hill, A. Briggs, Dr. D. B. Newell, Dr. A. R. Hight Walker, and Dr. F. Tavazza
National Institute of Standards and Technology, Gaithersburg, MD 20899, United States

Dr. H. Berger
École Polytechnique Fédérale de Lausanne (EPFL), Institut de Physique des Nanostructures, CH-1015 Lausanne, Switzerland

Metallic transition metal dichalcogenides like tantalum diselenide (TaSe$_2$) exhibit exciting behaviors at low temperatures, including the emergence of charge density wave (CDW) states. In this work, density functional theory (DFT) is used to investigate how structural, electronic, and Raman spectral properties of the CDW configuration change as a function of thickness. Such findings highlight the influence of dimensionality change (from 2D to 3D) and van der Waals (vdW) interactions on the system properties. The vdW effect is most strongly present in bulk TaSe$_2$ in the spectral range 165 cm$^{-1}$ to 215 cm$^{-1}$. The phonons seen in the experimental Raman spectra are compared with the results calculated from the DFT models as a function of temperature and layer number. The matching of data and calculations substantiates the model's description of the CDW structural formation as a function of thickness, which is shown in depth for 1L through 6L systems. These results highlight the importance of understanding interlayer interactions, which are pervasive in many quantum phenomena involving two-dimensional confinement.

## 1. Introduction

Layered two-dimensional (2D) materials, specifically transition metal dichalcogenides (TMDs), are a subject of ongoing interest because of the quantum phenomena that emerge as a result of reduced dimensionality.[1-4] These materials also display many interesting quantum phase changes, including superconductivity, charge (CDW), and spin (SDW) density waves. Particularly, the CDW phase in various polymorphous of $TaSe_2$, $TaS_2$, $NbSe_2$, $VSe_2$, and $TiSe_2$ has inspired many groups to investigate this phase's potential for applications in quantum information science.[5,6] Early investigations surrounding CDW phase transitions mainly were focused on bulk TMDs, with such transitions attributed to multiple mechanisms such as Fermi surface nesting,[7] saddle point singularities,[8] and electron-phonon interactions.[9-13] However, applications for quantum devices often require devices to be constructed from few-layer material to maximize the tunability of the material's properties.[14] Since van der Waals (vdW) interactions and reduced dimensionality contribute heavily to layer-dependent properties, comparisons should be made between monolayer (1L), few-layer systems and bulk, to learn more about how these effects contribute to the structural formation, and more specifically, to the formation of CDW states.[15-20]

Various works have been published, which, in their own merit, make observations or calculations that directly involve the effect of vdW interactions and reduced dimensionality on the CDW phase.[19, 21-26] Though many of those studies explored CDWs in $TaSe_2$ and related materials,[19, 21, 27-30] specific details on the layer-by-layer evolution of atomic structures and the formation of CDW phases are lacking. Some examples of structural studies include work by Ryu *et al.*, where they discussed the unique formation of triangular structures during the CDW phase transition in the 2D limits of $1T-TaSe_2$.[19] Another work reports the formation of a unidirectional (striped) CDW phase in $NbSe_2$.[30] Though some of the details of CDW phase transitions in the 2D limit are known, the intricacies of how the atomic structures and phonon modes evolve as a result of vdW interactions still require further investigations.

In this work, the focus will be primarily on thickness effects, corresponding to vdW interactions along *z* (the direction normal to the layers) becoming more prominent as the number of layers in a material increases. In materials that can be exfoliated, covalent, ionic, and metallic bonding act primarily only within each layer. As vdW interactions are much weaker than any of the aforementioned bonding types, their effect is negligible within the layer,[19] but significant along the direction normal to the layers, where no other bonding exists. Therefore, vdW interactions have no effect on the properties of monolayers (1L), i.e., for the purely two-dimensional (2D) cases. However, they become significant for bulk systems (3D case). When only a few layers are present, the in-between regime is not either completely 2D or 3D, so its behavior is unknown *a priori*, even with the bulk and 1L cases having been studied, and is therefore of extreme interest. Additionally, because of the reduced bonding, a higher degree of local disorder has been observed at finite temperatures in the 1L case than for thicker samples, leading to a substantial fluctuation in charge-ordered states that occasionally prevent the formation of long-range, coherent CDW states in the 2D limit.[21] Such deviations are not as resilient in bulk systems since atoms have increased interactions with neighboring layers, thereby restricting any atoms' small oscillations around their ideal positions. Again, the crossover point between monolayer and bulk behavior is not generally known with respect to this phenomenon.

The material of focus in this work is 2H-TaSe$_2$, which exhibits an incommensurate (IC-) CDW phase between 122 K and 90 K and a commensurate CDW phase below 90 K.[31,32] This material was chosen because it has already been used to develop Hall devices, optoelectronics, and similar applications.[33-36] Additionally, since the superconducting phase transition of 2H-TaSe$_2$ occurs at 0.2 K, the corresponding phase does not coexist with the CDW phase, making it easier to interpret calculations and experimental data. To date, CDW formation in 1L and few-layer 2H-TaSe$_2$ remains a subject of debate, warranting an elaborate investigation like the one presented here.[11, 37-43] To understand the impact of the vdW interactions on the CDW phases, several multilayer structures (up to six layers) have been investigated to explain the structural formation of the CDW phase. Knowledge of the structural formation of CDW phases will be crucial for devices whose functionality depends on activating specific phase transitions in few-layer 2H-TaSe$_2$ or similar materials.

**2. Regime of Reduced Dimensionality (1L-4L)**

To begin the analysis, it's important to reiterate that the influence of vdW interactions increases as the thickness of the sample (i.e., the number of layers) increases. For instance, in the two extreme cases of 1L and bulk systems, the vdW interaction, a layer-dependent effect, is important in the bulk case, as it is responsible for holding the layers together, while their effect is negligible in the case of a 1L system, due to a lack of adjacent neighbors with which to interact. The critical question is: how does the increased influence of vdW interactions affect the structural formation of CDW phases? If it does, the next logical question is: how many layers must a system have to exhibit signs of a crossover point to bulk-like behavior? The DFT models presented herein include vdW interactions and, therefore, investigate such a rivalry between the 1L and bulk limit. Our choice of material, 2H-TaSe$_2$, facilitates such a task since the CDW structures of 1L and bulk are quite different. Thus, transitioning from one to the other would require substantial changes that can be easily tracked.

A structural comparison was first made between 1L and bulk 2H-TaSe$_2$. The reason for doing so was to establish expected structural formation and electronic behaviors at the two extreme limits. The first step in determining the initial structures for this study was to relax the atomic position and lattice vectors of the bulk 2H-TaSe$_2$ unit cell (point group D$_{6h}$). The optimized lattice constants ($a_{TaSe2}$ = 0.339 nm and $c_{TaSe2}$ = 1.22 nm) agreed within the range of values from previous computational studies.[17, 20, 44, 45] Relaxed structural parameters, such as lattice constants, bond length, and intralayer distance, have been tabulated in Table S1. Starting with this relaxed bulk structure, a 1L (point group D$_{3h}$) and bulk supercell with 9-unit cells (3 × 3 × 1) were constructed. The trigonal prismatic structure of 2H-TaSe$_2$ can be found in Figure S1. Varying the electronic temperature (defined as σ) in discrete steps (see Supporting Information for details on the methodology), the evolution of the structure as a function of temperature, and the formation of the CDW state is modeled.

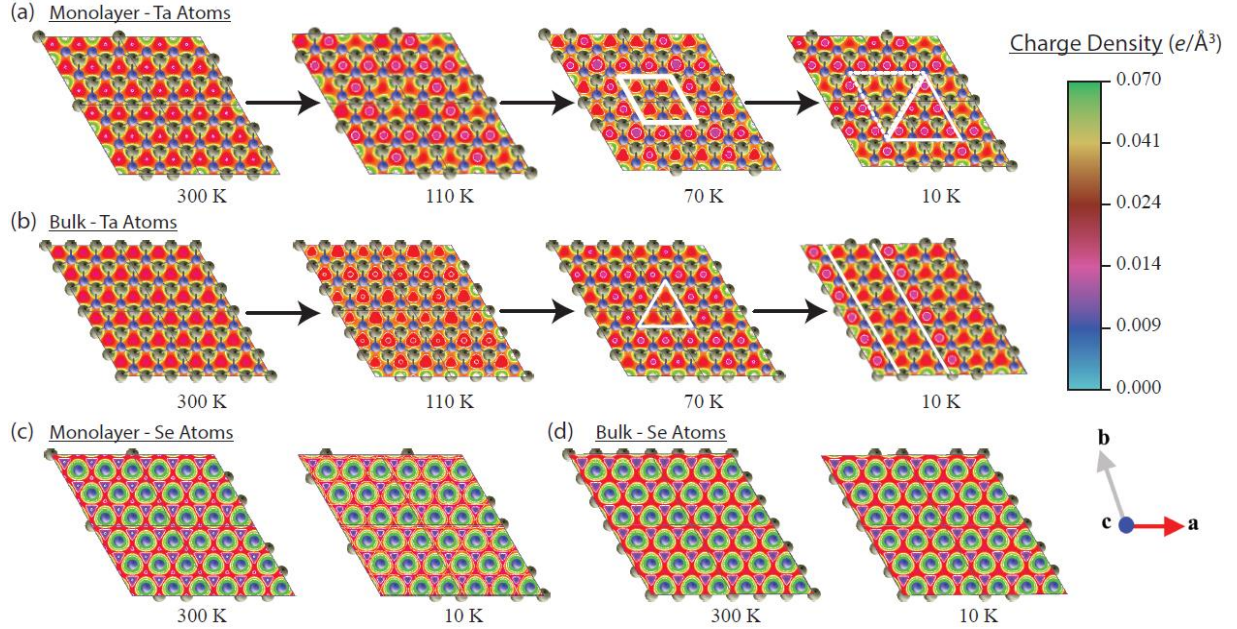

**Figure 1.** Evolution of the electron densities as temperature decreases. (a) 1L case, the atomic plane containing the Ta atoms. The formation of a rhombus-like structure occurs near 70 K that decomposes into a structure with a triangular appearance at 10 K, indicating the localization of charges. The solid and dotted white triangles mark such triangular formations. (b) In the bulk case, the plane containing the Ta atoms shows a triangular formation first (70 K), then decomposes into a striped lattice at 10 K. (c) No significant changes are seen, as a function of T, for the Se electronic densities, neither in the 1L nor in the (d) bulk case.

Electron densities are shown at key temperature steps in **Figure 1** for both 1L and bulk TaSe$_2$. Results for 300 K, 110 K, 70 K, and 10 K are shown for Ta atoms, whereas only the initial and final structural configurations (300 K and 10 K) are shown for the Se atoms. This reduced picture for Se atoms was shown because no significant changes are seen between high and low-temperature states in either the 1L or bulk cases, indicating that the Ta atoms contribute the most in terms of atomic rearrangement during the CDW transition. Despite not contributing heavily to atomic rearrangements, the Se will still have an important role in the interpretation of DFT predictions discussed later. Figure 1 (a) and (b), the temperatures of 110 K and 70 K, show structures in the IC-CDW and C-CDW phase, respectively, with the green and blue corresponding to higher and lower charge density areas, respectively. The complete structural evolution of the 1L and bulk systems as a function of temperature and the systems' transition into the CDW phase is shown in greater detail (10 K increments) in Figure S2 and Figure S3. From that detailed evolution, it is apparent that for temperatures less than 30 K (below the CDW transition), the formation of a triangular structure occurs in the 1L case instead of a striped structure formation as seen in the bulk case.

Additional details to note for this comparison of the 1L and bulk structures include one prominent similarity and three significant differences. For the former, a lack of charge density modulation is observed around the Se atoms; *i.e.*, the Ta atoms dominate the behavior. The first difference is that the Ta atoms' displacement starts at 175 K for the 1L but at a lower temperature (140 K) for the bulk case, as seen in Figure S3. Second and more importantly, a rhombus-like region of greater charge forms in the 1L case, and although both cases exhibit this greater charge at 70 K, the bulk case exhibits it in the form of triangular-like regions. Thirdly, at 10 K, when the C-CDW phase has reached its equilibrium structure, the 1L exhibits two adjacent triangular regions whereby one region has an accumulation of charge, and the adjacent region has charge depleted. Recall that in the bulk case, rather than forming triangular regions of charge (or lack of charge), a striped configuration forms whereby a periodic alternation between charge accumulation and depletion defines the bounds of those stripes.[45]

This formation of either triangular or striped structures for 1L or bulk systems, respectively, gives information about structural order. Incidentally, the structural rearrangements for the 1L case appear to be more complex as the triangular-like regions are being formed, a consequence of not having long-range order along the *z*-axis normally attributable to vdW interactions. The bulk case, being consistent with previous experimental observations of the CDW phase in NbSe$_2$,[30] continuously forms into a striped structure with decreasing temperature, indicating the presence of CDW long-range order.

Generally, other than the formation of triangular or striped structures at low temperatures, the charge distribution has a spherical symmetry very close to that of the Ta atom. The results confirm that this system is metallic, that the charge modulations completely depend on the Ta atomic displacement, and that vdW interactions, manifesting in the presence of neighboring layers in the DFT model, contribute significantly to predicted charge modulation differences between the 1L and bulk cases during the CDW phase transitions. The next natural question to ask concerns whether or not the presence of interacting, neighboring layers affects these ground-state structural formations and whether there is a coherent layer dependence that can link the two extreme cases of 1L and bulk.

Understanding how vdW interactions, via layer number, affect the structural formation of CDW phases requires knowledge from cases intermediate between 1L and bulk. Using DFT, we investigated the 2L, 3L, 4L, 5L, and 6L cases, with all cases starting from the same type of super-cells as built for 1L and bulk, and with all cases being examined as σ is reduced from 300 K to 10 K. All results shown below are for the structures identified as a ground state at 10 K. Additional tables and information about the ground state energy calculations are shown in the Supporting Information.

Figure 2 summarizes the first portion of these DFT findings. The 2L ground state structure, shown in Figure 2 (a), shows a clear symmetry when comparing the top and bottom layers, graphically represented by a vertical ellipsis (see Figure 2 (c)). This symmetry is equivalent to a 180° rotation. Finding similar structural features in the top and bottom layer supports this configuration being the ground state, not just a local minimum, as there are no physical reasons for the top and bottom to be different. More interestingly, no significant differences are found between the 2L structural arrangement on each atomic plane and the 1L structure at 10 K (Figure S2). This lack of structural change between the 1L and 2L cases most likely indicates that vdW interaction has only slightly increased in importance in the 2L material. The reduced dimensionality still dictates the electronic and ionic configurations. Other results of these calculations, like interlayer distances and bond lengths, are reported in the first two sections of the Supporting Information.

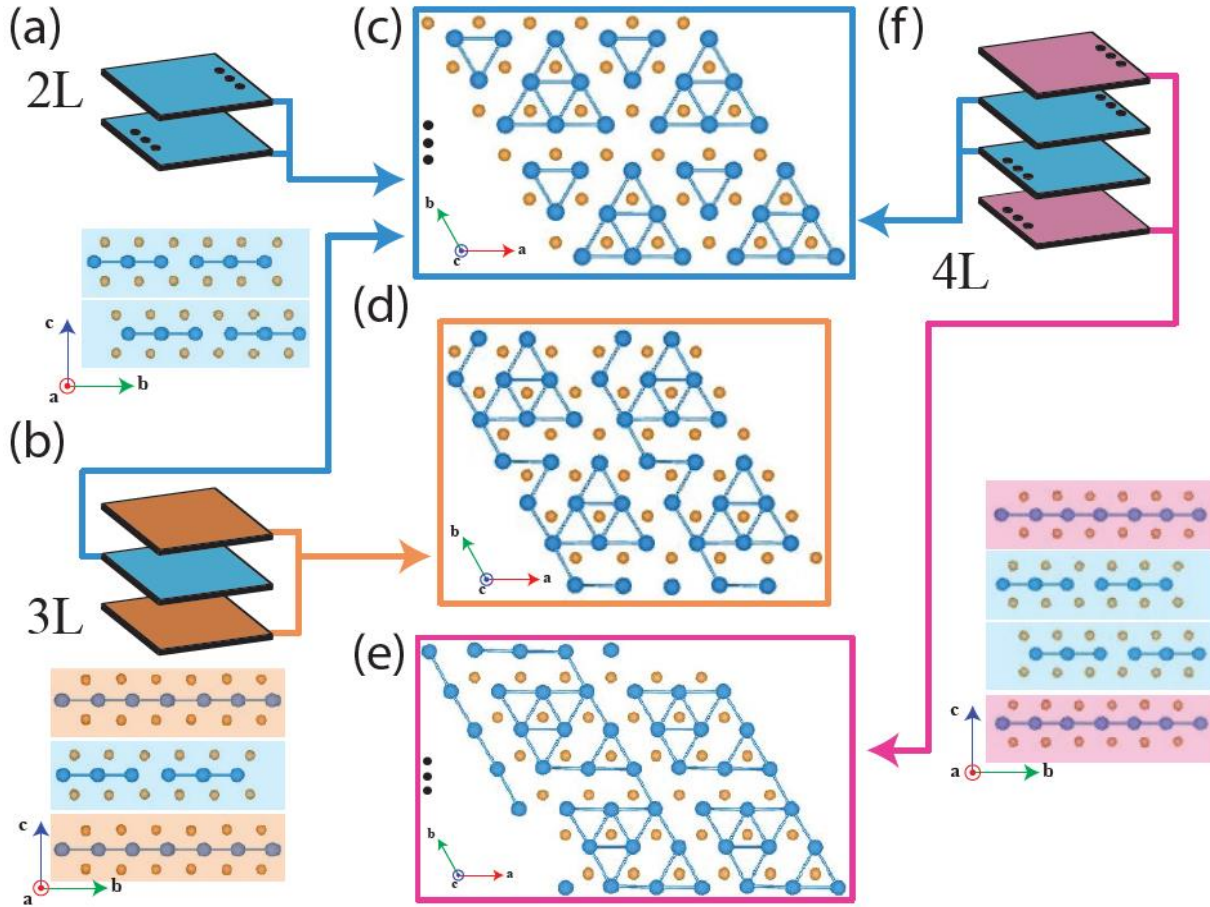

**Figure 2.** Calculated atomic structures are graphically represented for the (a) 2L and (b) 3L system with σ = 10 K. For the 2L system, whose side-view is seen in (a), the atomic arrangement is for the top and bottom is shown in (c), with a vertical ellipsis indicating layer orientation. For the 3L system in (b), each of the two distinct layer structures is shown, with the unitary middle layer depicted in (c) and the identical top and bottom layers shown in (d). The layer (pink) shown in (e) represents the top and bottom layers of the 4L system. (f) The side view of the calculated atomic structure for the 4L system is shown with the same vertical ellipsis indicating the layer orientation, as expected by the symmetry of the system (that is, the pink layers are rotated 180° with respect to each other). The same holds true for the inner two layers of the 4L system (blue), shown in (c).

In the 3L case, shown in Figure 2 (b), a departure from the usual triangular structure is observed. The middle layer continues to maintain its triangular structure, seen in Figure 2 (c) (blue), but the identical top and bottom layers, represented by an orange color in Figure 2 (d), begin to change in such a way that the smaller triangular features in the 1L case are no longer part of the structure. Instead, each of the larger triangular features becomes connected by a new periodic pattern reminiscent of a striped feature with missing components.

The 4L case is shown in Figure 2 (f), (c), and (e), with the side view in (f) and the two types of layers in (c) and (e), represented by the colors blue and pink, respectively. The middle two layers in blue retain an identical appearance to the 2L case. However, the outer two layers in pink continue the gradual transformation from the outer layers of the 3L case. The appearance of striped features becomes more evident in this case, and one should note that this phenomenon has only occurred with outer layers thus far.

**3. Regime of Meaningful Influence of van der Waals Interactions**

In Figure 3, the 5L case is shown, and this case marks a subtle but more crucial change in the trends that have been seen in the structures so far. Though the outer two layers continue to bear a resemblance to the 1L system with its triangular structure, the middle layers distinctly adopt an appearance incrementally closer to the bulk case – i.e., the formation of clearly striped structures. One intuitive way to picture this argument is that, as one goes from 1L to 6L, the first deviation away from the triangular structure of the central most layer (or layers for an even number) occurs at 5L. This is a gradual step towards having the internal layers of a thicker-layered system eventually match the bulk case, in the limit of tens of layers. A more quantifiable way of describing this transition is through counting the Ta bonds needed to form one structure or the other. For instance, to transform the triangular (1L) structure to a striped (bulk) structure (for a single plane so as to match the images), 22 Ta bonds must be added inside the supercell.[45] This number does not change until the 5L case, at which point 16 Ta bonds must be added to match the striped structure. It will be shown that the 6L case also requires its central-most layers to form 16 additional Ta bonds to match the bulk case.

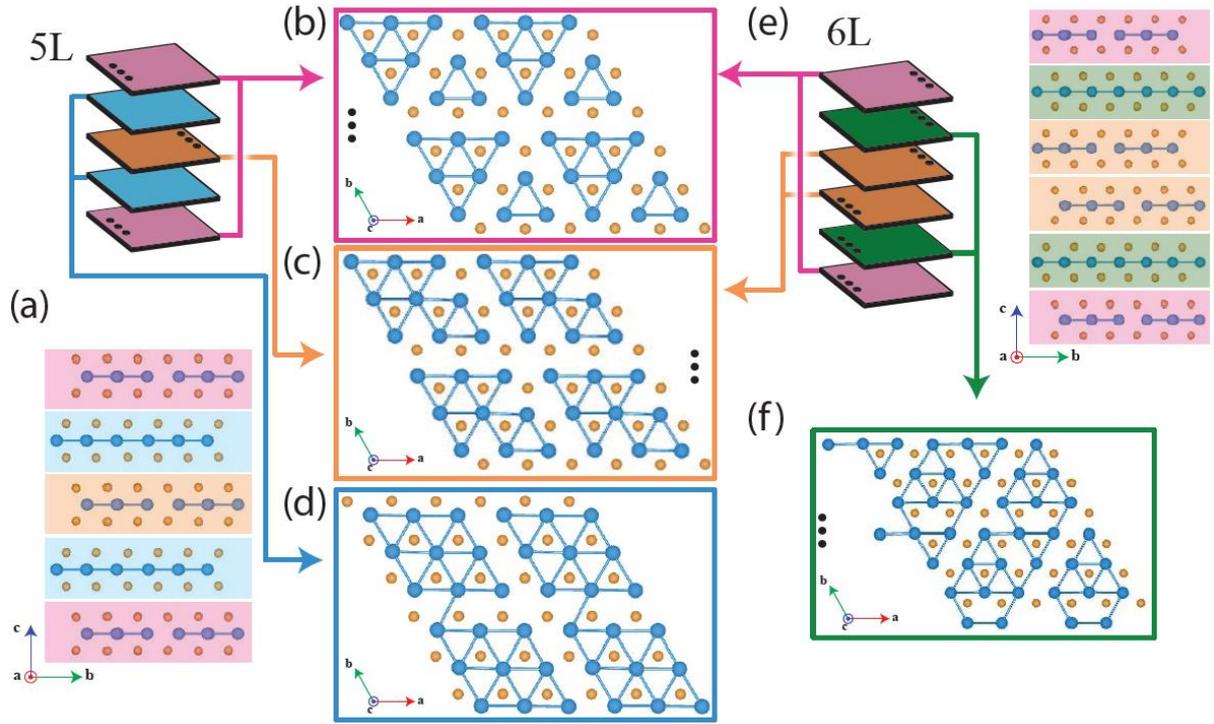

**Figure 3.** Calculated atomic structures are graphically represented for the 5L and 6L systems with σ = 10 K. The side-view for the 5L system is shown in (a), and each layer is color-coded to indicate the inherent symmetry in the system. The top and bottom layers (pink) are shown in (b), whereas the middle layer (orange) is shown in (c), and the two remaining inner layers (blue) are shown in (d). The side-view for the 6L system is shown in (e). An ellipsis serves as a point of reference to indicate a 180° rotational symmetry. The top and bottom layers (pink) are shown in (b). (f) The exact representation holds true for the second and fifth layers (green), as well as the (c) center two layers (orange).

The final case of 6L is also shown in Figure 3. What becomes evident in this case is that, more than in 5L, the material's internal structure tends towards stripe-like arrangements, as in the bulk case. Specifically, all four inner layers (Figure 5 (c) and (f)) begin to manifest a striped structure with a symmetry seen in lower even-numbered cases. The outermost layers, in pink, appear to revert to a triangular structure like that seen in the 2L and 3L cases. This is reasonable, as the overall strength of the vdW interaction is still limited (only 6 layers), and it is at its lowest on the outermost layers. Therefore, more than 6 layers are needed for bulk-like behavior, similarly with findings for other properties,[46-48], but 5 or 6 layers are enough to identify the general mechanism for such a transition and roughly estimate a crossover point between reduced dimensionality or vdW interactions having the stronger influence on the CDW structure.

As the energy landscape is particularly tricky to explore for this material, given that many local minima may be present at similar energies, we also investigated related ideal structures as a reference. Using the same supercell as in the previously discussed simulations and all thicknesses under examination, we utilized as initial states the CDW-ordered triangular and striped configurations found for the limiting 1L and bulk cases and relaxed them while discretely reducing σ. The comparison between the energies of these ideal, limiting structures (triangular and striped) and those found modeling the CDW transition directly reveals whether the modeling got trapped in a local minimum as well as enables the identification of the crossover point between 1L-like and bulk-like behavior.

More specifically, to accurately quantify this crossover point, which may also be regarded as a transition from the 2D to the 3D regime, two imposed structures have been considered while performing ground state energy calculations: striped, triangular, for 1L to bulk structures. In each of these cases, the initial structure, be it striped or triangular, is applied to all layers for each thickness. Thus, the final energies of the optimized structures are compared to the energies of the relaxed configurations discussed earlier in the paper to provide insight into the most stable structures (*i.e.*, those with the lowest energy ordering). The comparison of the energies per atom is shown in Fig. 4 as a function of reciprocal layer number so that the bulk system could be represented without breaking the axes. This plotting approach also allows straightforward determination of the crossover point from the intersection of the energy curves for the imposed-triangular and imposed-striped configurations.

The DFT calculations revealed that a system with forced striped layers is energetically unfavorable compared to a system that has been forced to exhibit only triangular structure up through 6L, at least. The energy difference between the forced striped and triangular structures decreases as the number of layers increases. At some critical point approaching the bulk case, shown in the inset of Fig. 4, the energetic favorability shifts from the triangular structure to the striped structure. The crossover point occurs at approximately 9L. Therefore, for this material in CDW configuration, 9 layers is a reasonable estimate for the crossover from the 2D to the 3D regime.

Compared to the forced cases, our system is always either energetically favorable or matching the most energetically favorable case, indicating that the simulations did not get trapped in some local minima. Specifically, for 1L and 2L, our system relaxed to a perfect triangular configuration, therefore matching the forced-triangular results. Similarly, as bulk, our system relaxed in a striped configuration, therefore matching the forced-striped case. For all mid-range thicknesses (3L-6L), our system always has the lowest energy, indicating that "mixed" configurations (some layers with configuration tending towards stripes, some similar to triangular) are the most convenient ones. This proves that, for these thicknesses, the effect of the vdW interactions is starting to become significant, but not completely there yet.

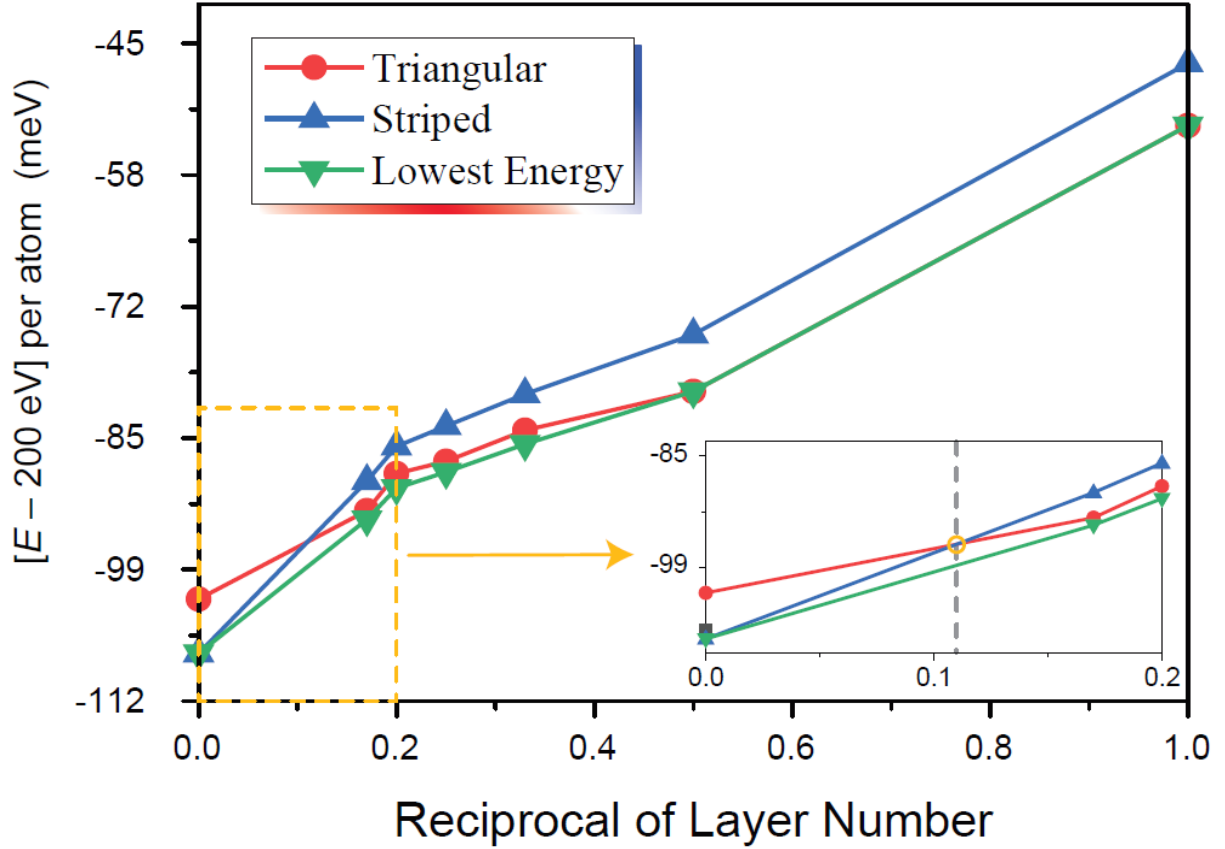

**Figure 4.** To more accurately quantify the crossover between the 2D and 3D regimes, three different structure types were imposed on the ground state energy calculations. In the case where every layer is forced to exhibit a triangular structure (red) from 1L to bulk, one obtains a trend for the energy per atom as a function of reciprocal layer number. By repeating these calculations for the case where every layer is forced to exhibit a striped structure (blue), one may compare the two latter cases and find that energetic favorability shifts between the two structure types around 9L (inset). When each layer takes on random atomic position perturbations (most representative of the experimental system), the trend takes on all lowest energies (green).

## 4. Experimental Results

Thus far, it has been demonstrated with DFT calculations that the presence of additional layers, and, therefore, a more significant magnitude of vdW interactions, leads to a gradual yet well-defined transition between having CDW phases represented by triangular versus striped structures. To validate this model, it was used to predict a set of CDW Raman modes for the 1L case, with the intent of making a comparison with experimental data. A similar comparison between Raman results obtained from this DFT model and experimental data has been detailed for the bulk case in recent work.[22] For frequencies greater than 170

cm$^{-1}$ and a temperature of 5 K (shaded region of Figure 5 (a)), experimental Raman spectra show CDW modes in the 1L, 2L, and bulk cases.

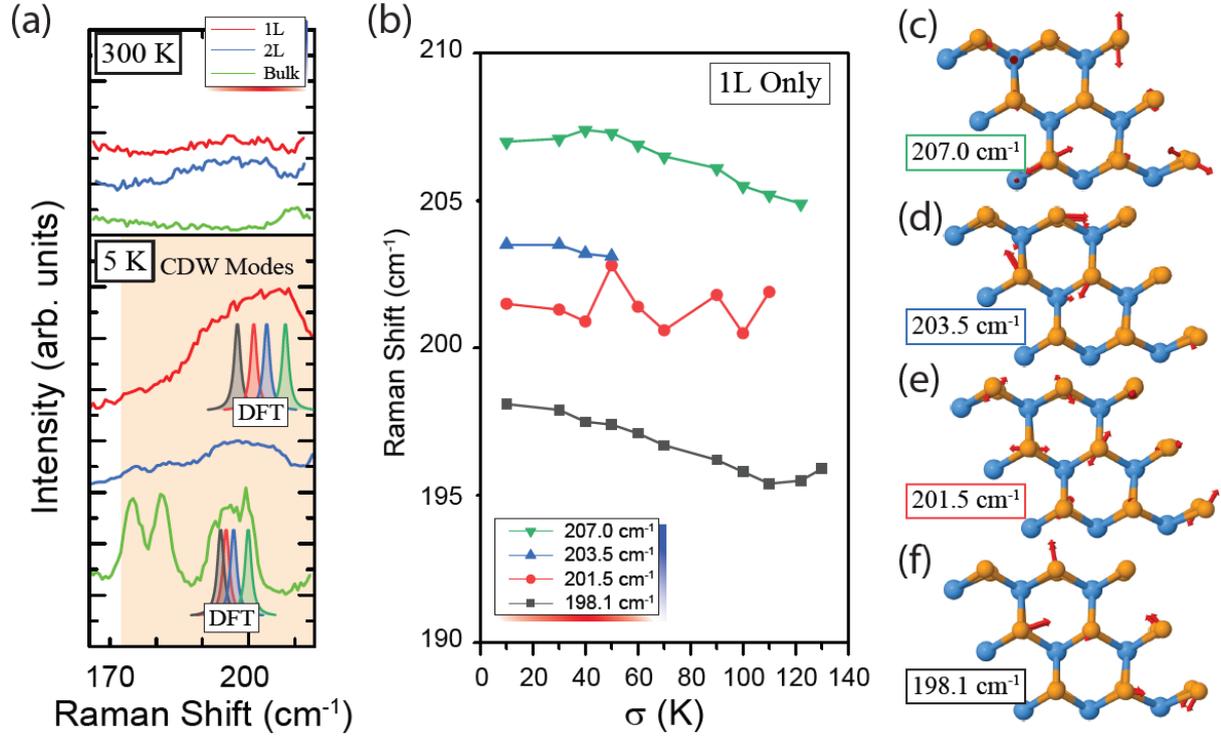

**Figure 5.** (a) Experimentally observed Raman spectra for the range between 170 cm$^{-1}$ and 215 cm$^{-1}$. Four illustrated peaks are drawn into the region containing the predicted CDW modes for the 1L case. (b) All four Raman-active, DFT-calculated CDW modes in the 1L case are plotted as a function of σ. Corresponding illustrations of the lattice modes are shown in (c)-(f) in order of increasing frequency, determined at the lowest temperature. Only the mode in (e) emerges in the C-CDW phase, whereas all others appear in the IC-CDW phase.

For the 1L case, four Raman-active modes in the CDW phase emerged from the DFT calculations. They are indicated by Gaussian peaks in Figure 5 (a). The experimental spectrum displays a single, broad peak roughly between 185 cm$^{-1}$ and 215 cm$^{-1}$, that consumes all four DFT peaks, therefore validating the DFT model. Though four CDW modes are seen in the bulk spectra as well, those peaks are redshifted with respect to those in the 1L and do not exhibit the same vibrational behavior as the ones described below.[45]

In addition to identifying modes, DFT allows one to investigate them as a function of temperature and in terms of vibration directions. The four predicted Raman-active CDW modes are plotted as a function of σ in Figure 5 (b), and their vibration directions are shown in Figure 5 (c) - (f). These results indicate that each mode's vibration and evolution are different, with three of them appear in the IC-CDW phase and one emerging in the C-CDW phase.

The 1L mode at 203.5 cm$^{-1}$, whose vibration is shown in Figure 5 (d), only appears at temperatures below 60 K, which is well within the C-CDW phase. It is a circular mode and does not shift in frequency as the temperature decreases. This circular mode involves Ta atoms vibrating in the counterclockwise direction, whereas both layers of Se atoms vibrate in the clockwise direction. This circular mode was also seen in the bulk case.[45] Additional temperature-dependent details for all four 1L modes are shown in section five of the Supporting Information.

The three remaining 1L modes emerge in the IC-CDW phase, characterized by a crystal symmetry containing an orthorhombic structure.[30, 49, 50] Two of them undergo a slight blueshift as the temperature decreases, and all of them have different vibration patterns (see Figure 5 (c), (e), and (f)): (i) a stretching bond mode (198.1 cm$^{-1}$); (ii) a breathing mode (201.5 cm$^{-1}$); and (iii) a triangular mode (207 cm$^{-1}$). The stretching mode (i) emerges around 130 K and involves vibrations roughly along the Se-Ta bond direction. The Ta atoms' vibrations are negligible compared to those of the Se atoms in this mode, with each layer of Se atoms vibrating in the same direction. The breathing mode (ii) emerges in the 1L spectra due to the breaking of translational symmetry along the *z*-axis,[51] which is consistent with the fact that no breathing modes were found in the bulk within a similar range of frequencies. According to the DFT calculations, the top and bottom layers of Se atoms vibrate in the opposite direction, with Ta vibrations being negligible. This breathing mode also depends on *intralayer* coupling and the atomic displacement of Ta atoms. The final CDW mode is the triangular mode (iii) (see Figure 5 (c)), where Se atoms' vibrations form two opposite triangles. The vibration directions of the Se and Ta atoms resemble a general breathing mode, with the top and bottom layers of Se vibrating in opposite directions, and the Ta atoms' vibrations are negligible. All in all, the predicted CDW modes in the 1L case were verified by the experimental data, giving support to the total DFT model whereby the atomic structure of the CDW phase goes from being triangular in the few-layer limit to becoming a striped structure in the bulk limit.

## 5. Conclusion

In this work, DFT was used to probe the effects of vdW interactions on the atomic structure of CDW phases in 2H-TaSe$_2$. From 1L through 6L, each system was investigated, resulting in the observation that short-range structural order, associated with the triangular structures of the 1L case, gradually transformed into the well-known striped structures seen in the bulk case. A defining thickness in this transformation was the 5L system, where the CDW phase structure most noticeably departs from having a triangular structure in its middle layers to a more striped-like configuration. The crossover from 2D to the 3D regime was estimated to occur for a 9L slab. To grant additional support to the model, we used it to predict four Raman CDW modes for 1L that were not well-covered in the literature. Excellent matches were found with the experimental results. Overall, the methods underlying this work can be extended to other 2D materials and explain the effects of vdW interactions on both their optical properties and quantum phase transitions.

## 6. Experimental Section/Methods

Calculations were carried out using DFT as implemented in the PWSCF code.[52-54] Within the local-density approximations (LDA), the Perdew-Zunger (PZ) exchange and correlation functionals were employed for the geometry optimizations and phonon calculations.[55] A LDA exchange-correlation functional was used because it yielded a better description of the optical properties of the material than does the general-gradient approximation (GGA).[56] Norm-conserving pseudopotentials were utilized for describing the interactions between core and valence electrons.[57, 58]

All the DFT calculations were performed at a thermal temperature of 0 K, but to model the temperature-dependent formation of the CDW states, the *electronic* temperature σ was modulated. This modulation can be done by tuning the smearing factor, a parameter which describes the Fermi-Dirac distribution, to qualitatively assess the effect of temperature on the phonon properties of the system and the Kohn anomaly.[4, 45, 59-62] We validated the approach of modeling real temperature effects with electronic temperature variations by computing the lattice expansion as a function of temperature and comparing it with experimental results.

Starting with the relaxed 1L structure, we constructed a supercell with 9 unit cells ($3 \times 3 \times 1$) and used a sufficiently large vacuum (20 Å) in the vertical direction to neglect any interaction between neighboring supercells. The kinetic energy cutoff of the plane-wave expansion is taken as 520 eV. All of the geometric structures were fully relaxed until the force on each atom was less than 0.002 eV Å$^{-1}$, and the energy-convergence criterion was $1 \times 10^{-7}$ eV. For the unit cell and supercell structure relaxation, a $16 \times 16 \times 16$ and $3 \times 3 \times 8$ *k*-point grid was used, respectively. For the phonon calculations, we used a $4 \times 4 \times 4$ uniform *q*-grid for the unit cell and only performed gamma point phonon calculations for the supercell.

For complementary experimental data, mechanically exfoliated, single crystals of 2H-TaSe$_2$ on Si/SiO$_2$ substrates (300 nm oxide layer) were prepared. During Raman spectrum acquisition, a 515 nm laser excitation was used at a sample temperature of both 5 K and 300 K. The scattered light was collected through a triple-grating spectrometer to enable low-frequency (down to approximately 10 cm$^{-1}$) measurements.

**References**

[1]     M. Chhowalla, D. Jena, and, H. Zhang, *Nat. Rev. Mater.*, **2016**, *1*, 16052.

[2]     A, J. Mannix, B. Kiraly, M. C. Hersam, N. P. Guisinger, *Nat. Rev. Chem.,* **2017**, *1*, 0014.

[3]     K. S. Novoselov, A. K. Geim, S. Morozov, D. Jiang, M. I. Katsnelson, I. Grigorieva, S. Dubonos, A. A. Firsov, *Nature*, **2005**, *438*, 197.

[4]     K. S. Novoselov, A. K. Geim, S. Morozov, D. Jiang, Y. Zhang, S. V. Dubonos, I. Grigorieva, and A. A. Firsov, *Science,* **2004**, *306*, 666-669.

[5]     H. Fröhlich, *Proc. R. Soc. Lond. A,* **1954**, *223*, 296-305.

[6]     R. E. Thorne, *Phys. Today*, **1996**, *49*, 42-48.

[7]     J. Wilson, F. Di Salvo, and S. Mahajan, *Phys. Rev. Lett.* **1974,** *32*, 882.

[8]     T. Rice and G. Scott, *Phys. Rev. Lett.* **1975,** *35*, 120.

[9]     M. Johannes and I. Mazin, *Phys. Rev. B* **2008,** *77*, 165135.

[10]    M. Johannes, I. Mazin, and C. Howells, *Phys. Rev. B* **2006,** *73*, 205102.

[11]    W. McMillan, *Phys. Rev. B* **1976,** *14*, 1496.

[12]    A. C. Neto, *Phys. Rev. Lett.* **2001,** *86*, 4382.

[13]    J. Laverock, D. Newby Jr, E. Abreu, R. Averitt, K. Smith, R. P. Singh, G. Balakrishnan, J. Adell, and T. Balasubramanian, *Phys. Rev. B* **2013**, *88*, 035108.

[14]    S. Kang, D. Lee, J. Kim, A. Capasso, H. S. Kang, J-W. Park, C-H. Lee, G-H. Lee, *2D Mater.,* **2020**, *7*, 022003.

[15]    R. Bianco, I. Errea, L. Monacelli, M. Calandra, and F. Mauri, *Nano Lett.* **2019,** *19*, 3098-3103.

[16]    G. Froehlicher, E. Lorchat, F. Fernique, C. Joshi, A. Molina-Sánchez, L. Wirtz, S. Berciaud, *Nano Lett.,* **2015**, *15*, 6481-6489.

[17]    Y. Ge, A. Y. Liu, *Phys. Rev. B* **2012,** *86*, 104101.

[18]    P. Hajiyev, C. Cong, C. Qiu, T. Yu, *Sci. Rep.* **2013,** *3*, 2593.

[19]    H. Ryu, Y. Chen, H. Kim, H.-Z. Tsai, S. Tang, J. Jiang, F. Liou, S. Kahn, C. Jia, A. A. Omrani, *Nano Lett*. **2018,** *18*, 689-694.

[20]    J.-A. Yan, M. A. D. Cruz, B. Cook, K. Varga, *Sci. Rep.* **2015,** *5*, 16646.

[21] X. Xi, L. Zhao, Z. Wang, H. Berger, L. Forró, J. Shan, K. F. Mak, *Nat. Nanotechnol.* **2015,** *10*, 765.

[22] D. Lin, S. Li, J. Wen, H. Berger, L. Forro, H. Zhou, S. Jia, T. Taniguchi, K. Watanabe, X. Xi, M. S. Bahramy, *Nat. Commun.* **2020**, *11*, 1-9.

[23] M. Yoshida, R. Suzuki, Y. Zhang, M. Nakano, Y. Iwasa, *Sci. Adv.* **2015,** *1*, e1500606.

[24] J. Yang, W. Wang, Y. Liu, H. Du, W. Ning, G. Zheng, C. Jin, Y. Han, N. Wang, Z. Yang, *Appl. Phys. Lett.* **2014,** *105*, 063109.

[25] P. Chen, Y.-H. Chan, X.-Y. Fang, Y. Zhang, M.-Y. Chou, S.-K. Mo, Z. Hussain, A.-V. Fedorov, T.-C. Chiang, *Nat. Commun.* **2015,** *6*, 1-5.

[26] J. Feng, D. Biswas, A. Rajan, M. D. Watson, F. Mazzola, O. J. Clark, K. Underwood, I. Markovic, M. McLaren, A. Hunter, *Nano Lett.* **2018,** *18*, 4493-4499.

[27] L. Bawden, S. Cooil, F. Mazzola, J. Riley, L. Collins-McIntyre, V. Sunko, K. Hunvik, M. Leandersson, C. Polley, T. Balasubramanian, *Nat. Commun.* **2016,** *7*, 11711.

[28] M. M. Ugeda, A. J. Bradley, Y. Zhang, S. Onishi, Y. Chen, W. Ruan, C. Ojeda-Aristizabal, H. Ryu, M. T. Edmonds, H.-Z. Tsai, *Nat. Phys.* **2016,** *12*, 92.

[29] X. Xi, Z. Wang, W. Zhao, J.-H. Park, K. T. Law, H. Berger, L. Forró, J. Shan, K. F. Mak, *Nat. Phys.* **2016,** *12*, 139.

[30] A. Soumyanarayanan, M. M. Yee, Y. He, J. Van Wezel, D. J. Rahn, K. Rossnagel, E. W. Hudson, M. R. Norman, J. E. Hoffman, *Proc. Natl. Acad. Sci.* **2013,** *110*, 1623-1627.

[31] D. Moncton, J. Axe, F. DiSalvo, *Phys. Rev. Lett.,* **1975**, *34*, 734.

[32] S. Sugai, K. Murase, *Phys. Rev. B*, **1982**, *25*, 2418.

[33] M. Mahajan, S. Kallatt, M. Dandu, N. Sharma, S. Gupta, K. Majumdar, *Commun. Phys.* **2019,** *2*, 1-9.

[34] A. T. Neal, Y. Du, H. Liu, P. D. Ye, *ACS Nano* **2014,** *8*, 9137-9142.

[35] Y. Shan, L. Wu, Y. Liao, J. Tang, X. Dai, Y. Xiang, *J. Mater. Chem. C* **2019,** *7*, 3811-3816.

[36]     H. Wang, Y. Chen, C. Zhu, X. Wang, H. Zhang, S. H. Tsang, H. Li, J. Lin, T. Yu, Z. Liu, *Adv. Funct. Mater.* **2020,** *30*, 2001903.

[37]     R. N. Bhatt, W. McMillan, *Phys. Rev. B*, **1975**, *12*, 2042.

[38]     S. Borisenko, A. Kordyuk, A. Yaresko, V.Zabolotnyy, D. Inosov, R. Schuster, B. Büchner, R. Weber, R. Follath, L. Patthey, *Phys. Rev. Lett.,* **2008**, *100*, 196402.

[39]     J. A. Holy, M. V. Klein, W. McMillan, S. Meyer, *Phys. Rev. Lett.,* **1976**, *37*, 1145.

[40]     K. Rossnagel, E. Rotenberg, H. Koh, N. Smith, L. Kipp, *Phys. Rev. B*, **2005**, *72*, 121103.

[41]     K. Rossnagel, N. Smith, *Phys. Rev. B*, **2007**, *76*, 073102.

[42]     J. Tsang, J. Smith Jr, M. Shafer, *Solid, State, Commun.,* **1978**, *27*, 145-149.

[43]     J. A. Wilson, *Phys. Rev. B*, **1978**, *17*, 3880.

[44]     S. Chowdhury, J. R. Simpson, T. Einstein, A. R. Hight Walker, *Phys. Rev. Mater.* **2019,** *3*, 084004.

[45]     H. M. Hill, S. Chowdhury, J. R. Simpson, A. F. Rigosi, D. B. Newell, H. Berger, F. Tavazza, A. R. Hight Walker, *Phys. Rev. B*, **2019**, *99*, 174110.

[46]     J. Zhu, J. Wu, Y. Sun, J. Huang, Y. Xia, H. Wang H. Wang, Y. Wang, Q. Yi, G. Zou, *RSC Adv.* **2016**, *6*, 110604-9.

[47]     Y. Cai, G. Zhang, Y. W. Zhang, *Sci. Rep.* **2014,** *4*, 6677.

[48]     R. Beams, L. G. Cançado, S. Krylyuk, I. Kalish, B. Kalanyan, A. K. Singh, K. Choudhary, A. Bruma, P. M. Vora, F. Tavazza, A. V. Davydov, *ACS Nano*, **2016,** *10,* 9626-36.

[49]     J. van Wezel, J. *Euro. Phys. Lett.* **2011,** *96*, 67011.

[50]     J. van Wezel, P. Littlewood, *Physics* **2010,** *3*, 87.

[51]     D. L. Duong, G. Ryu, A. Hoyer, C. Lin, M. Burghard, K. Kern, *ACS Nano* **2017,** *11*, 1034-1040.

[52]     P. Hohenberg, W. Kohn, *Phys. Rev.,* **1964**, *136*, B864.

[53]    W. Kohn, L. J. Sham, *Phys. Rev.,* **1965**, *140*, A1133.

[54]    P. Giannozzi, S. Baroni, N.Bonini, M. Calandra, R. Car, C. Cavazzoni, D. Ceresoli, G. L. Chiarotti, M. Cococcioni, I. Dabo, *J. Phys. Condens. Mater.,* **2009**, *21*, 395502.

[55]    J. P. Perdew, A. Zunger, *Phys. Rev. B*, **1981**, *23*, 5048.

[56]    J. P. Perdew, J. A. Chevary, S. H. Vosko, K. A. Jackson, M. R. Pederson, D. J. Singh, C. Fiolhais, *Phys. Rev. B*, **1992**, *46*, 6671.

[57]    M. Fuchs, M. Scheffler, *Comp. Phys. Commun.,* **1999**, *119*, 67-98.

[58]    N. Troullier, J. L. Martins, *Phys. Rev. B*, **1991**, *43*, 1993.

[59]    M. Calandra, I. Mazin, F. Mauri, *Phys. Rev. B*, **2009**, *80*, 241108.

[60]    D. L. Duong, M. Burghard, J. C. Schön, *Phys. Rev. B*, **2015**, *92*, 245131.

[61]    W. Kohn, *Phys. Rev. Lett.,* **1959**, *2*, 393.

[62]    F. Weber, S. Rosenkranz, J-P. Castellan, R. Osborn, R. Hott, R. Heid, K-P. Bohnen, T. Egami, A. Said, D. Reznik, *Phys. Rev. Lett.,* **2011**, *107*, 107403.

# Supporting Information

**Influence of Dimensionality on the Charge Density Wave Phase of 2H-TaSe$_2$**

*Sugata Chowdhury[*], Albert F. Rigosi, Heather M. Hill, Andrew Briggs, David B. Newell, Helmuth Berger, Angela R. Hight Walker, and Francesca Tavazza*

§I. Crystal structure of bulk and monolayer 2H-TaSe$_2$

§II. Structural parameters

§III. Evolution of structure with temperature for monolayer 2H-TaSe$_2$

§IV. Evolution of structure with temperature for bulk 2H-TaSe$_2$

§V. Temperature-dependent evolution of the four 1L modes

§VI. Other layer-dependent properties

**§I. Crystal structure of bulk and monolayer 2H-TaSe$_2$:**

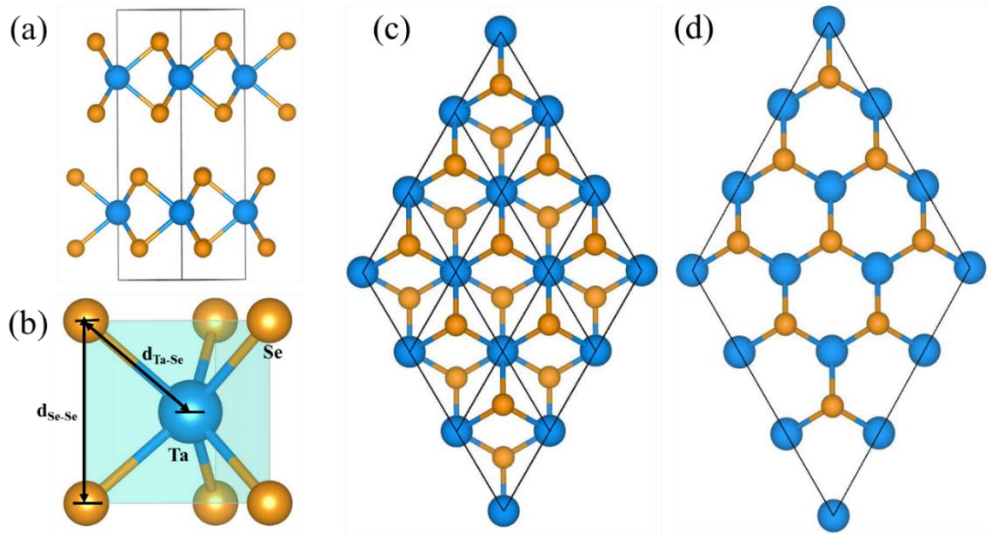

**Fig S1:** Representative structures of 2H-TaSe$_2$ monolayer (1L) and bulk. Ta and Se atoms are shown as blue and yellow atoms, respectively. (a) The bulk crystal structure of 2H-TaSe2. (b) The trigonal prismatic coordination of 2H-TaSe$_2$ phase. To investigate the CDW structure, we constructed a $3 \times 3 \times 1$ supercell. In (c) and (d), we have shown the $3 \times 3 \times 1$ supercell of bulk and 1L 2H-TaSe$_2$, respectively.

## §II. Structural parameters:

|  | $a_0$ (Å) | $c_0$ (Å) | $d_{Se-Se}$ (Å) | $d_{Se-Ta}$ (Å) |
|---|---|---|---|---|
| Bulk Exp[1] | 3.43 | 12.71±0.04 | 3.35±0.02 | 2.59±0.01 |
| Bulk-DFT[2] | 3.37 | 12.34 | 3.37 | 2.53 |
| 1L-DFT | 3.38 | - | 3.50 | 2.61 |
| 2L-DFT | 3.37 | - | 3.49 | 2.60 |
| 3L-DFT | 3.39 | - | 3.47 | 2.59 |
| 4L-DFT | 3.39 | - | 3.45 | 2.59 |
| 5L-DFT | 3.39 | - | 3.43 | 2.58 |
| 6L-DFT | 3.39 | - | 3.41 | 2.58 |

**Table S1.** Lattice parameters, bond length (Se-Ta), and the distance between Se-Se atomic layers of the unit cell of 2H-TaSe$_2$ are calculated using DFT.

## §III. Evolution of structure with temperature for monolayer 2H-TaSe$_2$.:

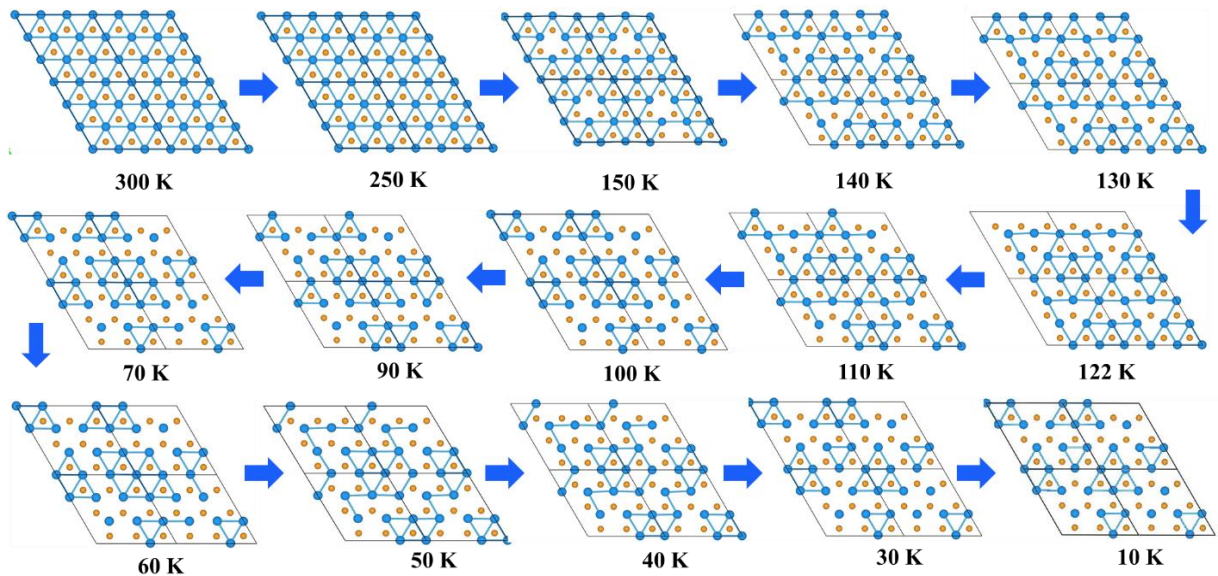

**Fig S2.** DFT-calculated atomic rearrangement driven by the formation of the CDW phase for 1L 2H-TaSe$_2$, viewed from above. Ta-Ta bonds (blue lines) are drawn for a Ta atom separation of about 0.340 nm.

## §IV. Evolution of structure with temperature for bulk 2H-TaSe$_2$:

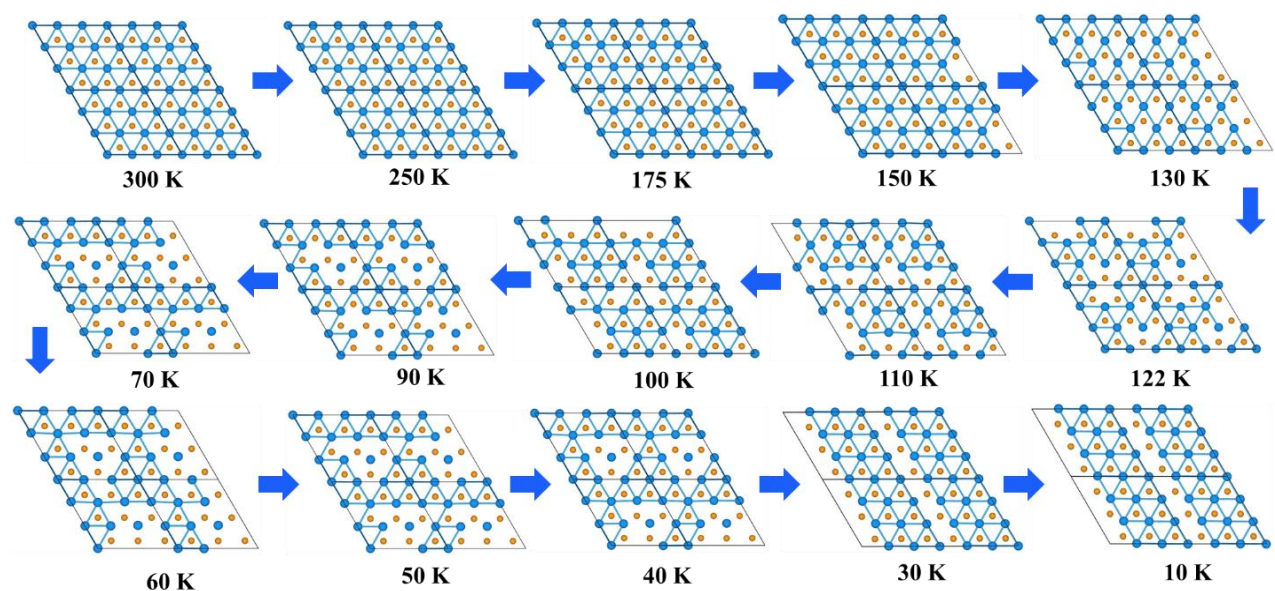

**Fig S3.** DFT-calculated atomic rearrangement driven by the formation of the CDW phase for bulk 2H-TaSe$_2$, viewed from above. Ta-Ta bonds (blue lines) are drawn for a Ta atom separation of about 0.340 nm.

## §V. Temperature-dependent evolution of the four 1L modes

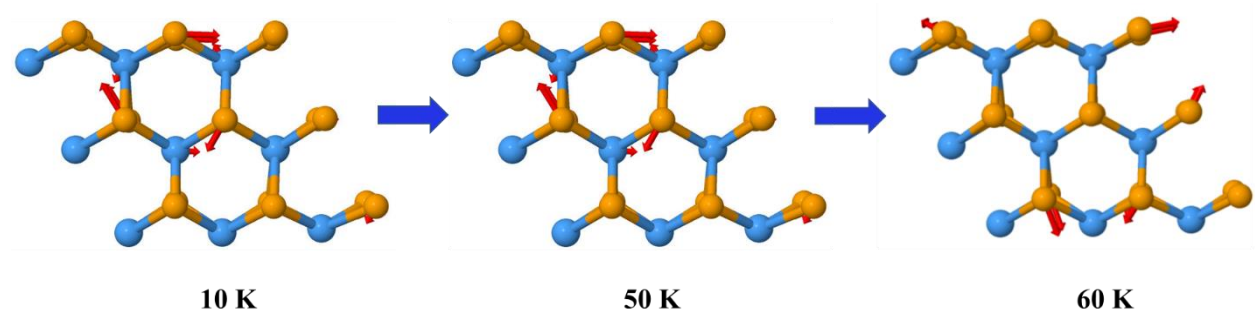

**Fig S7.** A circular mode at 203.5 cm$^{-1}$ appears in the C-CDW phase below 60 K. Se atoms' vibrations are exhibited in a clockwise direction, whereas those of the Ta atoms are exhibited in a counterclockwise direction.

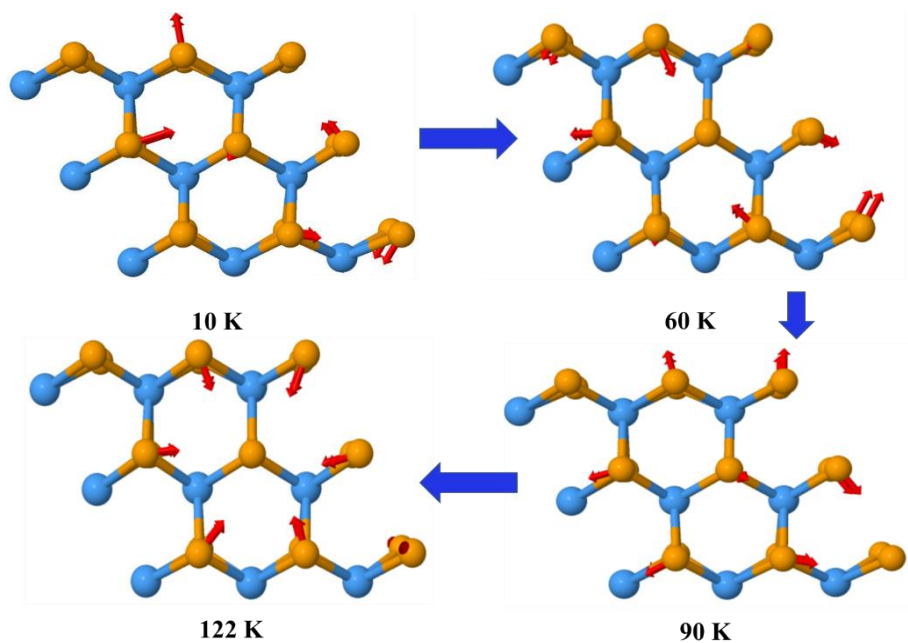

**Fig S9.** A stretching mode at 198.1 cm$^{-1}$ appears in the IC-CDW phase below 122 K.

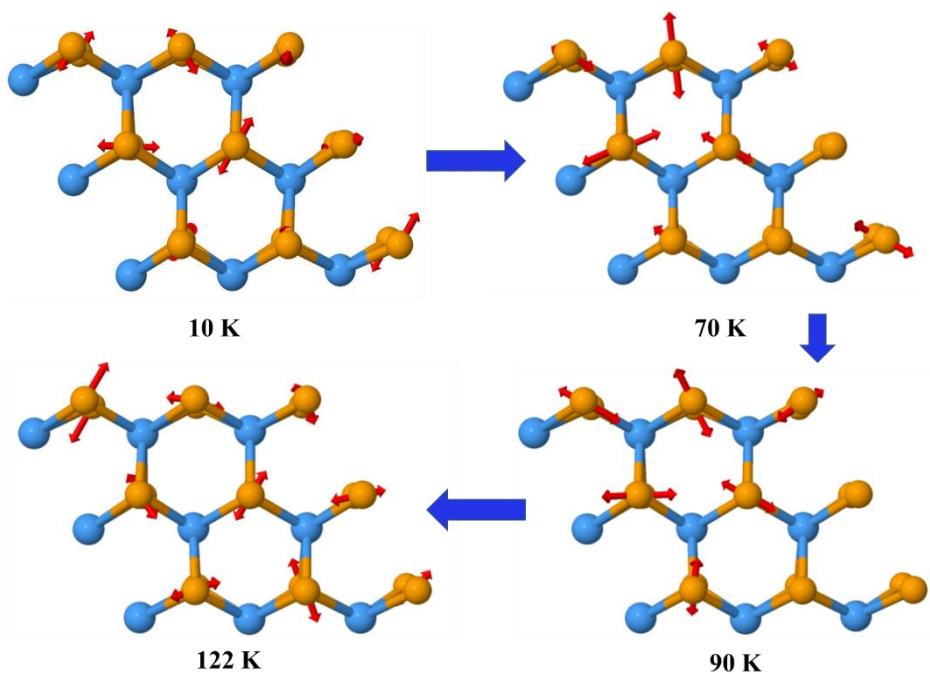

**Fig S8.** A breathing mode at 201.5 cm$^{-1}$ appears in the IC-CDW phase below 122 K.

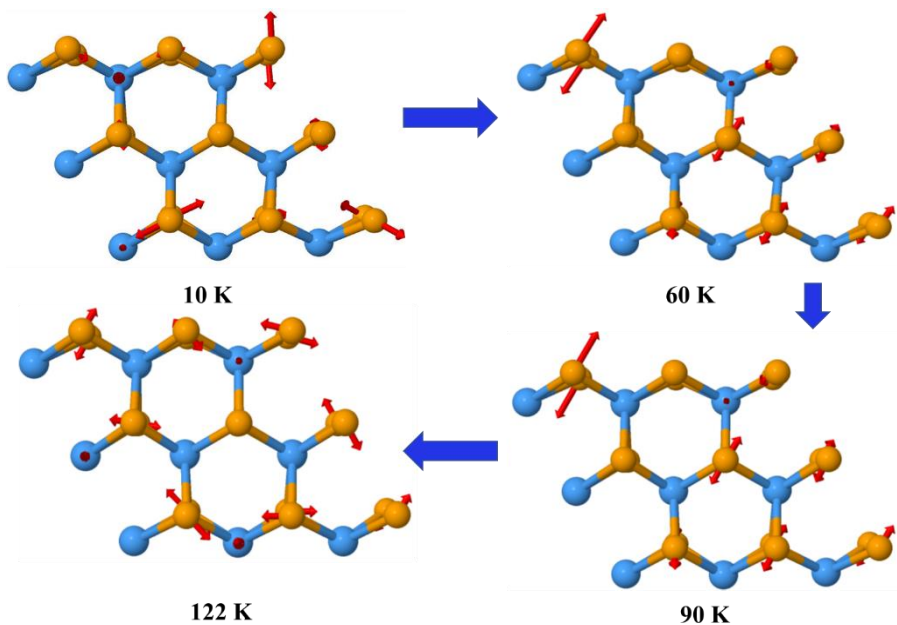

**Fig S9.** A Triangular mode at 207.0 cm$^{-1}$ appears in the IC-CDW phase below 122 K.

## §VI. Other layer-dependent properties

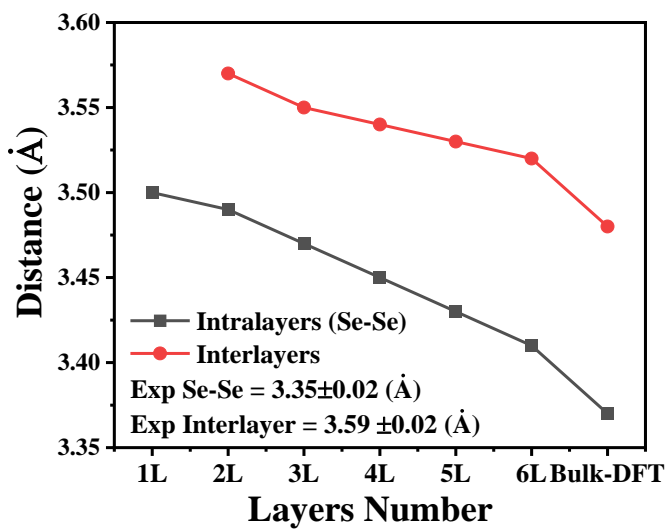

**Fig S10.** The plot below shows both the interlayer spacing and intralayer Se-Se bond lengths as a function of layer-number. These values occurred for the ground state calculations of each system and are compared with experimental results.[1]

| Case | # Atoms | Our Ground State (Ry) | Hexagonal (Ry) | Triangle (Ry) | Stripes (Ry) |
|---|---|---|---|---|---|
| 1L | 27 | -436.6878284 | -436.6857197 | -436.6878284 | -436.6750233 |
| 2L | 54 | -873.4848246 | -873.4821631 | -873.4848246 | -873.4614378 |
| 3L | 81 | -1310.259769 | -1310.248813 | -1310.25121 | -1310.228917 |
| 4L | 108 | -1747.035897 | -1747.009374 | -1747.027003 | -1746.998097 |
| 5L | 135 | -2183.811245 | -2183.751586 | -2183.796355 | -2183.768461 |
| 6L | 162 | -2620.612133 | -2620.513384 | -2620.601293 | -2620.564843 |
| Bulk | 54 |  | -873.5879935 | -873.5701258 | -873.5923611 |

| Case | # Atoms | Hex. (Ryd per atom) | Tri. (Ryd per atom) | Striped (Ryd per atom) | Our Result (Ryd per atom) |
|---|---|---|---|---|---|
| 1L | 27 | -16.17355 | -16.17362 | -16.17315 | -16.17362 |
| 2L | 54 | -16.17560 | -16.17564 | -16.17521 | -16.17564 |
| 3L | 81 | -16.17591 | -16.17594 | -16.17567 | -16.17605 |
| 4L | 108 | -16.17601 | -16.17618 | -16.17591 | -16.17626 |
| 5L | 135 | -16.17594 | -16.17627 | -16.17606 | -16.17638 |
| 6L | 162 | -16.17601 | -16.17655 | -16.17633 | -16.17662 |
| Bulk | 54 | -16.17756 | -16.17722 | -16.17764 |  |

**Table S2.** Ground state energies for various cases, with considerations given to specific sublattices present in the CDW phase for 1L-like cases (manifesting as triangular sublattices) and bulk-like cases (manifesting as striped sublattices).

**Supporting Information References:**

1. B. E. Brown, D. J. Beerntsen, *Acta Crystallogr.* **1965,** *18*, 31-36.
2. H. M. Hill, S. Chowdhury, J. R. Simpson, A. F. Rigosi, D. B. Newell, H. Berger, F. Tavazza, A. R. Hight Walker, *Phys. Rev. B*, **2019**, *99*, 174110.